\documentstyle[preprint,aps]{revtex}

\begin{document}
\draft
\thispagestyle{empty}
\begin{titlepage}
\preprint{CAS-HEP-T-95-11/003, OHSTPY-HEP-T-95-023, hep-ph/9511327}
\title{ CP Violation, Fermion Masses and Mixings 
 in a Predictive SUSY $SO(10)\times \Delta (48) \times U(1)$ Model with 
Small $\tan \beta $}
\author{K.C. Chou$^{1}$ and Y.L. Wu$^{2}$\footnote{ 
supported in part by Department of Energy Grant\# DOE/ER/01545-662}}
\address{ $^{1}$Chinese Academy of Sciences, Beijing 100864,  China  
 \\ $^{2}$Department of Physics, \ Ohio State  University \\ Columbus, 
 Ohio 43210,\ U.S.A. }
\date{Rapid Communication, Phys. Rev. D {\bf 53}, No. 7 (1996)} 
\maketitle

\begin{abstract}
Fermion masses and mixing angles are studied in an
SUSY $SO(10)\times \Delta (48)\times U(1)$ model with small $\tan\beta$.
Thirteen parameters involving masses 
and mixing angles in the quark and charged lepton 
sector are successfully predicted by a single Yukawa coupling and three 
ratios of VEVs caused by necessary symmetry breaking. 
Ten relations among the low energy parameters have been found
with four of them free from renormalization modifications. 
They could be tested directly by low energy experiments.
\end{abstract}
\pacs{PACS numbers: 12.15.Ff, 11.30.Er, 12.10.Dm, 12.60.Jv}

\end{titlepage}

\newpage
\narrowtext
 
% text file
 
The standard model (SM) is a great success. Eighteen phenomenological 
parameters in the SM, which are introduced to describe all the low energy data, 
have  been extracted from various experiments although they are not yet
equally well known. Some of them have an accuracy of  better than $1\%$, 
but some others less than $10 \%$. To improve the accuracy for 
these parameters and understand them is a big challenge for 
particle physics. The mass spectrum and the mixing angles observed 
remind us that we are in a stage similar to that of  atomic 
spectroscopy before Balmer.
 Much effort has been made along this direction. 
The well-known examples are the
Fritzsch ansatz\cite{FRITZSCH} and Georgi-Jarlskog texture\cite{GJ}. 
 A general analysis and review of the 
previous studies on the texture structure was given by Raby 
in \cite{RABY}.  Recently, Babu, and Barr\cite{BB}, and Mohapatra\cite{BM}, 
and Shafi \cite{BS}, Hall and Raby\cite{HR}, Berezhiani\cite{BE}, 
Kaplan and Schmaltz\cite{KS}, Kusenko and Shrock \cite{KSH} constructed
some interesting models with texture zeros based on supersymmetric (SUSY)
SO(10).  Anderson,  
Dimopoulos, Hall, Raby,  and Starkman\cite{OPERATOR} presented a general 
operator analysis for the quark and charged lepton Yukawa coupling 
matrices with two zero textures `11' and `13'. 
The 13 observables in the quark and charged lepton
sector were found to be successfully fitted by only six parameters with large
$\tan \beta$. Along this direction, we have shown\cite{CHOUWU} that  the same
13 parameters can be successfully 
described,  in an SUSY $SO(10)\times \Delta(48)\times U(1)$ model
with large values of $\tan\beta \sim m_{t}/m_{b}$, by only five 
parameters with three of them determined  
by the symmetry breaking scales of U(1), SO(10), SU(5), and $SU(2)_{L}$.
Ten parameters in the neutrino sector could also be predicted, 
though not unique, with one additional parameter. 

   In this Rapid Communication, we shall present, 
based on the symmetry group SUSY $SO(10)\times \Delta(48)\times U(1)$, 
 an alternative model with small values of $\tan \beta \sim 1 $ which 
is of phenomenological interest in testing the Higgs 
sector in the minimum supersymmetric standard model (MSSM)
at   Colliders\cite{ELLIS}. 
The dihedral group $\Delta (48)$, 
a subgroup of SU(3), is taken as the family group.  
$U(1)$ is family-independent and is introduced to distinguish 
various fields which belong to the same 
representations of $SO(10)\times \Delta (48)$.
The irreducible representations of $\Delta (48)$ consisting of five triplets 
and three singlets  are found to be sufficient to build an interesting
texture structure for fermion mass matrices.  The 
symmetry $\Delta(48)\times U(1)$  naturally 
ensures the texture structure with zeros for Yukawa coupling matrices, 
while the coupling coefficients of the resulting interaction terms in the
superpotential are unconstrainted by this symmetry. 
To reduce the possible free parameters, 
the universality of coupling constants in the superpotential is assumed, i.e., 
all the coupling coefficients are assumed to be equal and have the same 
origins from perhaps a more fundamental theory. We know in  general that
universality of charges occurs only in the gauge interactions due to charge
conservation like the electric charge of different
particles. In the absence of strong interactions family symmetry could keep
the universality of weak interaction in 
a good approximation after breaking. In our
case there are so many heavy fermions above the grand unification theory 
(GUT) scale and their
interactions are taken to be universal in the GUT scale where family symmetries
have been broken. It can only be an ansatz at the
present moment where we do not know the answer governing  
 the behavior of nature above the GUT scale.
As the numerical predictions on the low energy parameters so found are 
very encouraging and interesting,  we believe that there must be a  
deeper reason that has to be found in the future.

 Choosing the structure of the physical vacuum carefully, 
the Yukawa coupling matrices which 
determine the masses and mixings of all quarks and leptons are given by
\begin{equation}
 \Gamma_{u}^{G} = \frac{2}{3}\lambda_{H} 
\left( \begin{array}{ccc} 
0  &  \frac{3}{2}z'_{u} \epsilon_{P}^{2} &   0   \\
\frac{3}{2}z_{u} \epsilon_{P}^{2} &  - 3 y_{u} 
\epsilon_{G}^{2} e^{i\phi}  
& -\frac{\sqrt{3}}{2}x_{u}\epsilon_{G}^{2}  \\
0  &  - \frac{\sqrt{3}}{2}x_{u}\epsilon_{G}^{2}  &  w_{u} 
\end{array} \right)
\end{equation}   
and
\begin{equation}
 \Gamma_{f}^{G} = \frac{2}{3}\lambda_{H} \frac{(-1)^{n+1}}{3^{n}} 
\left( \begin{array}{ccc} 
0  &  -\frac{3}{2}z'_{f} \epsilon_{P}^{2} &   0   \\
-\frac{3}{2}z_{f} \epsilon_{P}^{2} &  3 y_{f} 
\epsilon_{G}^{2} e^{i\phi}  
& -\frac{1}{2}x_{f}\epsilon_{G}^{2}  \\
0  &  -\frac{1}{2}x_{f}\epsilon_{G}^{2}  &  w_{f} 
\end{array} \right)
\end{equation}   
for $f=d,e$,  and 
\begin{equation}
\Gamma_{\nu}^{G} = \frac{2}{3}\lambda_{H}\frac{1}{5}\frac{(-1)^{n+1}}{15^n} 
\left( \begin{array}{ccc} 
0  &  -\frac{15}{2}z'_{\nu} \epsilon_{P}^{2} &   0   \\
-\frac{15}{2}z_{\nu} \epsilon_{P}^{2} &  15 y_{\nu} 
\epsilon_{G}^{2} e^{i\phi}  
& -\frac{1}{2}x_{\nu}\epsilon_{G}^{2}  \\
0  &  -\frac{1}{2}x_{\nu}\epsilon_{G}^{2}  &  w_{\nu} 
\end{array} \right)
\end{equation}   
for Dirac-type neutrino coupling,  where the integer $n$ reflects the possible
choice of heavy fermion fields above the GUT scale.  $n=4$ is found to be the
best choice in this set of models for 
a consistent prediction on top and charm quark masses.
This is because for $n > 4$, the resulting value of $\tan \beta$ becomes too
small, as a consequence, the predicted top quark mass will be below the present
experimental lower limit. For $ n < 4$, the values of $\tan \beta$ will become
larger, the resulting charm quark mass will be above the present upper bound. 
 $\lambda_{H}$ is an universal coupling constant expected to be of order one.
$\epsilon_{G}\equiv v_{5}/v_{10}$ and 
$\epsilon_{P}\equiv v_{5}/\bar{M}_{P}$  with $\bar{M}_{P}$, $v_{10}$, and 
$v_{5}$ being the vacuum expectation values (VEVs) 
for $U(1)\times \Delta(48)$, SO(10) and SU(5) 
symmetry breaking
respectively. $\phi$ is the physical CP phase 
arising from the VEVs. The assumption of maximum CP violation implies that 
$\phi = \pi/2$. 
$x_{f}$, $y_{f}$, $z_{f}$, and $w_{f}$ $(f = u, d, e, \nu)$ 
are the Clebsch factors of $SO(10)$ determined by the 
directions of symmetry breaking of the adjoints {\bf 45}'s. 
The three directions of symmetry breaking have been chosen as
 $<A_{X}>=v_{10}\  diag. (2,\ 2,\ 2,\ 2,\ 2)\otimes \tau_{2}$, 
$<A_{z}> =v_{5}\  diag. (-\frac{2}{3},\ -\frac{2}{3},\ -\frac{2}{3},\ 
-2,\ -2)\otimes \tau_{2}$, 
$<A_{u}>=v_{5}\  
diag. (\frac{2}{3},\ \frac{2}{3},\ \frac{2}{3},\ 
 \frac{1}{3},\  \frac{1}{3})\otimes \tau_{2}$.
The Clebsch factors associated with the symmetry breaking directions can be 
easily read off from the U(1) hypercharges of the adjoints {\bf 45}'s and the
related effective operators which are obtained when the symmetry 
$SO(10)\times \Delta (48)\times U(1)$ is broken and heavy fermion pairs 
are integrated out and decoupled: 
\begin{eqnarray} 
W_{33} & = & (\frac{2}{3}\lambda_{H})\  \frac{1}{2} 
\ 16_{3} \  \frac{\sqrt{3}}{\sqrt{1 + 2(\frac{v_{10}}{A_{X}})^{2(n+1)}}}\ 
 (\frac{v_{10}}{A_{X}})^{n+1}\ 10_{1}\ (\frac{v_{10}}{A_{X}})^{n+1} \ 
\frac{\sqrt{3}}{\sqrt{1 + 2(\frac{v_{10}}{A_{X}})^{2(n+1)}}}\ 16_{3}  
\nonumber \\
W_{32} & = & (\frac{2}{3}\lambda_{H}) \frac{\sqrt{3}}{2} 
\epsilon_{G}^{2}\ 16_{3} 
\frac{\sqrt{3}}{\sqrt{1 + 2(\frac{v_{10}}{A_{X}})^{2(n+1)}}}\ 
 (\frac{v_{10}}{A_{X}})^{n+1}\ 
(\frac{A_{z}}{v_{5}})(\frac{v_{10}}{A_{X}})\ 10_{1}\ 
(\frac{v_{10}}{A_{X}}) (\frac{A_{z}}{v_{5}}) 
 (\frac{v_{10}}{A_{X}})^{n+1}\ 16_{2}  \nonumber \\
W_{22} & = & (\frac{2}{3}\lambda_{H}) \frac{3}{2} \epsilon_{G}^{2}
\ 16_{2}\ (\frac{v_{10}}{A_{X}})^{n}\ 
(\frac{A_{u}}{v_{5}})(\frac{v_{10}}{A_{X}})\ 10_{1}\  (\frac{v_{10}}{A_{X}}) 
(\frac{A_{u}}{v_{5}})\   (\frac{v_{10}}{A_{X}})^{n}\ 16_{2}\  e^{i\phi}   \\
W_{12} & = & (\frac{2}{3}\lambda_{H}) \frac{3}{2}  \epsilon_{P}^{2} \  
16_{1}\  [(\frac{v_{10}}{A_{X}})^{n-3}\ 10_{1}\  
(\frac{v_{10}}{A_{X}})^{n-3} \nonumber \\
 & & + (\frac{v_{10}}{A_{X}})^{n}\ 
(\frac{A_{u}}{v_{5}})(\frac{v_{10}}{A_{X}})\ 10_{1}\  (\frac{v_{10}}{A_{X}}) 
(\frac{A_{z}}{v_{5}}) (\frac{v_{10}}{A_{X}})^{n+1} ]\ 16_{2} \nonumber
\end{eqnarray}
The factor $1/\sqrt{1 + 2(\frac{v_{10}}{A_{X}})^{2(n+1)}}$  
arising from the mixing, is equal to $1/\sqrt{3}$ for the up-type quark and 
almost  unity for other fermions due to suppression of large Clebsch 
factors in the second
term of the square root. The relative phase (or sign) between  the
two terms in the operator $W_{12}$ has been fixed.
The resulting Clebsch  factors are
$w_{u}=w_{d}=w_{e}=w_{\nu} =1$, 
$x_{u}= 5/9$, $x_{d}= 7/27$, $x_{e}=-1/3$, $x_{\nu} = 1/5$
$y_{u}=0$, $y_{d}=y_{e}/3=2/27$, $y_{\nu} = 4/45$,  
$z_{u}=1$, $z_{d}=z_{e}= -27$, $z_{\nu} = -15^3 = -3375$, 
$z'_u = 1-5/9 = 4/9$, $z'_d = z_d + 7/729 \simeq z_{d}$,
$z'_{e} = z_{e} - 1/81 \simeq z_{e}$,   
$z'_{\nu} = z_{\nu} + 1/15^{3} \simeq z_{\nu}$.  

 An adjoint {\bf 45} $A_{X}$ and a 16-dimensional representation 
Higgs field $\Phi$ ($\bar{\Phi}$) 
are needed for breaking SO(10) down to SU(5). 
Another two adjoint 45s $A_{z}$ and $A_{u}$ are needed to break SU(5) further 
down to the standard model 
$SU(3)_{c} \times SU_{L}(2) \times U(1)_{Y}$.  
From the Yukawa coupling matrices given above ,
the 13 parameters in the SM can be determined by only four parameters: a
universal coupling constant $\lambda_{H}$ and three ratios of the VEVs: 
$\epsilon_{G}=v_5/v_{10}$, $\epsilon_{P}=v_5/\bar{M}_P$ and 
$\tan \beta = v_2/v_1 $. 
In obtaining physical masses and mixings, 
renormalization group (RG) effects should be taken into account.
As most Yukawa couplings in the present model are much smaller than the top 
quark Yukawa coupling $\lambda_{t}^{G} \sim 1$, in a good approximation, we
will only keep top quark Yukawa coupling terms in the RG equations and 
neglect all other Yukawa coupling terms.
 The RG evolution  will be described by
three kinds of  scaling factors. 
$\eta_{F}$ ($F=U,D,E,N$)  and $R_{t}$
 arise from running the Yukawa parameters from the GUT scale 
down to the SUSY breaking scale $M_{S}$ which is chosen to be close to
the top quark mass, i.e., $M_{S} \simeq m_{t}\simeq 170$ GeV. 
They are defined by $\eta_{F}(M_{S}) = \prod_{i=1}^{3}
\left(\frac{\alpha_{i}(M_{G})}{\alpha_{i}(M_{S})}\right)^{c_{i}^{F}/2b_{i}}$ 
$(F=U, D, E, N)$ with $c_{i}^{U} = (\frac{13}{15}, 3, \frac{16}{3})$, 
$c_{i}^{D} = (\frac{7}{15}, 3, \frac{16}{3})$ ,  
$c_{i}^{E} = (\frac{27}{15}, 3, 0)$, $c_{i}^{N} = (\frac{9}{25}, 3, 0)$,   
$b_{i} = (\frac{33}{5}, 1, -3)$,  
and 
$ R_{t}^{-1} = exp[-\int_{\ln M_{S}}^{\ln M_{G}} 
(\frac{\lambda_{t}(t)}{4\pi})^{2} dt ]  
=[1 + (\lambda_{t}^{G})^{2} K_{t}]^{-1/12}$, 
where $K_{t} = \frac{3 I(M_{S})}{4\pi^{2}}$ with 
$I(M_{S}) = \int_{\ln M_{S}}^{\ln M_{G}} \eta_{U}^{2}(t) dt $. 
The numerical value for $I$ taken from Ref. \cite{FIX} is 113.8 
for $M_{S} \simeq m_{t} = 170$GeV. Other RG scaling factors are derived by
running Yukawa couplings below $M_{S}$. $m_{i}(m_{i}) = \eta_{i} \  
m_{i} (M_{S})$ for $(i = c,b )$ and 
 $m_{i}(1GeV) = \eta_{i}\  m_{i} (M_{S})$ for ($i = u,d,s$).
The physical top quark mass is given by 
$M_{t} = m_{t}(m_{t}) \left(1 +
\frac{4}{3}\frac{\alpha_{s}(m_{t})}{\pi}\right)$. 
Using the well-measured charged lepton masses 
$m_{e}= 0.511$ MeV, $m_{\mu} = 105.66$ MeV, and 
$m_{\tau} = 1.777$ GeV, we obtain four important RG scaling-independent
predictions:
\begin{eqnarray}
& & |V_{us}| = |{V_{us}}|_{G} \simeq 
3 \sqrt{\frac{m_{e}}{m_{\mu}}} \left( \frac{1 +
(\frac{16}{675}\frac{m_{\tau}}{m_{\mu}})^{2}}{1 + 9
\frac{m_{e}}{m_{\mu}}}\right)^{1/2} = {\bf 0.22} , \\
& & |\frac{V_{ub}}{V_{cb}}|= |\frac{V_{ub}}{V_{cb}}|_{G} \simeq 
(\frac{4}{15})^{2} \frac{m_{\tau}}{
m_{\mu}}\sqrt{\frac{m_{e}}{m_{\mu}}} = {\bf 0.083} , \\
& & |\frac{V_{td}}{V_{ts}}|= |\frac{V_{td}}{V_{ts}}|_{G} \simeq 
3 \sqrt{\frac{m_{e}}{m_{\mu}}} = {\bf 0.209 }, \\
& & \frac{m_{d}}{m_{s}} \left(1- \frac{m_{d}}{m_{s}} \right)^{-2} 
= 9 \frac{m_{e}}{m_{\mu}}
\left(1- \frac{m_{e}}{m_{\mu}} \right)^{-2} = {\bf 0.044}
\end{eqnarray}
and six RG scaling-dependent predictions
\begin{eqnarray}
& & |V_{cb}| = |{V_{cb}}|_{G} R_{t} \simeq 
\frac{15\sqrt{3}- 7}{15\sqrt{3}}\frac{5}{4\sqrt{3}}
 \frac{m_{\mu}}{m_{\tau}} R_{t}  = {\bf 0.0391}\ 
\left(\frac{0.80}{R_{t}^{-1}}\right), \\
& & m_{s}(1GeV) = \frac{1}{3} m_{\mu} \frac{\eta_{s}}{\eta_{\mu}} \eta_{D/E}
= {\bf 159.53}\  \left(\frac{\eta_{s}}{2.2}\right) \left(\frac{\eta_{D/E}}{2.1}
\right)\  MeV, \\
& & m_{b}(m_{b}) = m_{\tau} \frac{\eta_{b}}{\eta_{\tau}} \eta_{D/E} R_{t}^{-1} 
= {\bf 4.25} \  \left( \frac{\eta_{b}}{1.49}\right) 
\left( \frac{\eta_{D/E}}{2.04}\right) 
\left( \frac{R_{t}^{-1}}{0.80} \right) \  GeV \ , \\
& & m_{u}(1GeV) = \frac{5}{3}(\frac{4}{45})^{3} \frac{m_{e}}{m_{\mu}} \eta_{u} 
R_{t}^{3} m_{t} = {\bf 4.23}\ \left(\frac{\eta_{u}}{2.2}\right)
\left(\frac{0.80}{R_{t}^{-1}}\right)^{3} 
\left( \frac{m_{t}(m_{t})}{174 GeV}\right)\  MeV \ , \\ 
& & m_{c}(m_{c}) = \frac{25}{48} (\frac{m_{\mu}}{m_{\tau}})^{2} 
\eta_{c} R_{t}^{3} m_{t} = {\bf 1.25}\  \left(\frac{\eta_{c}}{2.0}\right)
\left(\frac{0.80}{R_{t}^{-1}}\right)^{3} 
\left( \frac{m_{t}(m_{t})}{174 GeV}\right)\ GeV ,  \\
& & m_{t}(m_{t}) = \frac{\eta_{U}}{\sqrt{K_{t}}} \sqrt{1 - R_{t}^{-12}}
\frac{v}{\sqrt{2}} \sin \beta = {\bf 174.9}\  \left( \frac{\sin \beta}{0.92}
\right)  \left( \frac{\eta_{U}}{3.33}\right) 
\left( \sqrt{ \frac{8.65}{K_{t}}}\right)  \left(
\frac{\sqrt{1-R_{t}^{-12}}}{0.965} \right) \  GeV   
\end{eqnarray}
where the miraculus numbers in  the above relations are due to
the Clebsch factors. The scaling factor $R_{t}$ or coupling 
$\lambda_{t}^{G} = \frac{1}{\sqrt{K_{t}}} \frac{\sqrt{1 -
R_{t}^{-12}}}{R_{t}^{-6}}$  is 
determined by the mass ratio of the bottom quark and $\tau$ lepton.
$\tan \beta$ is fixed by the $\tau$ lepton mass via 
$\cos \beta = \frac{m_{\tau} \sqrt{2}}{\eta_{E} \eta_{\tau} v 
\lambda_{\tau}^{G}} $.  

The above 10 relations are our main results which contain only low energy
observables. As an analogy to the Balmer series formula, 
these relations may be
considered as empirical at the present moment.  
They have been tested by the existing
experimental data to a good approximation and can be tested further directly
by more precise experiments in the future. 

In numerical predictions we take $\alpha^{-1}(M_Z) = 127.9 $, $s^{2}(M_Z) 
= 0.2319$, $M_Z = 91.187$ GeV, 
$\alpha_{1}^{-1}(m_t) = 58.59$, 
$\alpha_{2}^{-1}(m_t) = 30.02$ and 
$\alpha_{1}^{-1}(M_G) = \alpha_{2}^{-1}(M_G) = \alpha_{3}^{-1}(M_G)
 \simeq 24 $ with $M_{G} \sim 2 \times 10^{16}$ GeV.  For $\alpha_{s}(M_{Z}) = 
0.113$, the RG scaling factors have values ($\eta_{u,d,s}$, $\eta_{c}$,
$\eta_{b}$, $\eta_{e,\mu,\tau}$, $\eta_{U}$, $\eta_{D}/\eta_{E}\equiv
\eta_{D/E}$, $\eta_{E}$, $\eta_{N}$) = (2.20, 2.00, 1.49, 1.02, 3.33, 2.06,
1.58, 1.41).
The corresponding predictions on fermion masses and mixings thus obtained are 
found to be remarkable.  Our numerical
predictions for $\alpha_{s} (M_{Z}) = 0.113$ are given in table 1 with   four 
input parameters: three charged lepton masses and bottom quark mass
$m_{b}(m_{b}) = 4.25$GeV,  where 
$B_{K}$ and $f_{B} \sqrt{B}$ in table 1 are two important hadronic
parameters and extracted from $K^{0}-\bar{K}^{0}$ and  $B^{0}-\bar{B}^{0}$
mixing parameters $\varepsilon_{K}$ and $x_{d}$. $Re(\varepsilon'/\varepsilon)$
is the direct CP-violating parameter in kaon decays, where large 
uncertanties mainly arise from the hadronic matrix elements. $\alpha$, 
$\beta$ and $\gamma$ are three angles of the unitarity triangle in the 
Cabibbo-Kobayashi-Maskawa (CKM) matrix. 
$J_{CP}$ is the repahse-invariant CP-violating quantity. 

 It is amazing that nature has allowed us to 
make predictions on fermion masses and mixings 
in terms of a single Yukawa coupling constant and 
three ratios of the VEVs determined by the structure of the physical 
vacuum and  understand the low energy physics 
from the GUT scale physics. It has also suggested that nature favors 
 maximal spontaneous CP violation. 
A detailed analysis including the neutrino sector will be presented in 
a longer paper \cite{CW}. In comparison with the models 
with large $\tan \beta \sim m_{t}/m_{b}$, 
the present model has provided a consistent 
picture on the 13 parameters in the SM with better accuracy. Besides, 
ten relations involving fermion masses and CKM matrix elements
are obtained with four of them independent of the RG scaling effects.
The two types of the models corresponding to the 
large and low $\tan\beta$
might  be distinguished by testing the MSSM Higgs sector at Colliders as well
as by precisely measuring the ratio $|V_{ub}/V_{cb}|$ since this ratio does
not receive radiative corrections in both models.   
It is expected that more precise measurements from CP violation and various low
energy experiments in the near future could provide crucial tests on the 
ten realtions obtained in the present model. 
\\

{\bf ACKNOWLEDGEMENTS}:\   YLW would like to thank Professor 
S. Raby and Professor G. Steigman for useful discussions.
\\
\newpage

{\bf  Table 1.}  Output parameters and their predicted values 
with $\alpha_{s}(M_{Z}) = 0.113$ and input parameters: $m_{e} = 0.511$ eV, 
$m_{\mu} = 105.66$ MeV, $m_{\tau} = 1.777$ GeV, and $m_{b} = 4.25$ GeV.
\\

\begin{tabular}{|c|c|c|c|c|}   \hline
 Output parameters   &  Output values   &  Data\cite{DATA} & 
 Output para.   &  Output values    \\ \hline 
$M_{t}$\ [GeV]  &  182   &  $180 \pm 15 $  &  $J_{CP} = A^{2} 
\lambda^{6} \eta $ & $2.68 \times 10^{-5}$  \\
$m_{c}(m_{c})$\ [GeV]  &  1.27   & $1.27 \pm 0.05$  & 
 $\alpha$ & $86.28^{\circ}$ \\ 
$m_{u}$(1GeV)\ [MeV]  &  4.31   &  $4.75 \pm 1.65$ & 
$\beta$ & $22.11^{\circ}$ \\
$m_{s}$(1GeV)\ [MeV]  &  156.5  &  $165\pm 65$  &  
$\gamma$ & $71.61^{\circ}$  \\
$m_{d}$(1GeV) \ [MeV]  &  6.26 & $8.5 \pm 3.0$ & 
$\tan \beta = v_{2}/v_{1}$ & 2.33  \\
$|V_{us}|=\lambda $ & 0.22 & $0.221 \pm 0.003$ & 
$\epsilon_{G}=v_{5}/v_{10}$ &  $2.987 \times 10^{-1}$ \\
$\frac{|V_{ub}|}{|V_{cb}|} = \lambda \sqrt{\rho^{2} + \eta^{2}}$ & 0.083 & 
$0.08 \pm 0.03$ & 
$\epsilon_{P} =v_{5}/\bar{M}_{P}$  & $1.011 \times 10^{-2}$ \\
$\frac{|V_{td}|}{|V_{ts}|} = \lambda \sqrt{(1-\rho)^{2} + \eta^{2}}$ & 0.209 & 
$0.24 \pm 0.11$ & 
$\lambda_{t}^{G}$  & 1.30 \\
 $|V_{cb}|=A\lambda^{2}$ & 0.0393  &  $0.039 \pm 0.005\cite{VCB} $ & - & - \\
$B_{K}$ & 0.90 &  $0.82 \pm 0.10$\cite{BK,LATTICE} & - & - \\
$f_{B}\sqrt{B}$ [MeV] & 207  & $200 \pm 70\cite{FB,LATTICE} $ & - & -  \\
 $Re(\varepsilon'/\varepsilon) $ & $(1.4 \pm 1.0)\cdot 10^{-3}$ &  
$(1.5 \pm 0.8 )\cdot 10^{-3}$ & - & - 
  \\  \hline
\end{tabular}
\\
\\

\end{document}